\documentstyle[12pt]{article} 
\begin{document}
\title{On the Integrability and Chaotic behaviour of an
ecological model}
\author{M. P. Joy\thanks{email : joy@mrc.iisc.ernet.in} \\
Materials Research Centre, Indian Institute of Science\\
Bangalore - 560 012, India. }
\date{}
\maketitle
\begin{abstract}
A three species food chain model is studied analytically as
well as numerically. Integrability of the model is studied using
Painlev\'e analysis while chaotic behaviour is studied
using numerical techniques, such as calculation of Lyapunov
exponents, plotting the bifurcation diagram and phase
plots.  We correct and critically comment on the wrong
results reported recently on this ecological model,   in a
paper by Rai ([1995] ``Modelling ecological systems'', Int.
J. Bifurcation and Chaos {\bf 5}, 537-543).

\end{abstract}
\vspace{2cm}


Short Title  : Chaos in an ecological model

\newpage
\section{Introduction}
Several studies on chaos in continuous as well as discrete
ecological models are available in the literature [May,
1976; Gilpin, 1979; Schaffer, 1985; Schaffer \& Kot, 1985;
Hastings \& Powell, 1991].  Recently Rai [1995] has
presented some studies on an ecological model, where he has
used mainly Painlev\'e analysis for studying the
integrability and has given  results of some numerical calculations 
also. The calculations given in that paper is
completely wrong and the claims he is making based on these
wrong results are misleading and confusing.  Almost all
results presented in that paper have been published in
another paper by Rai et al. [1993].   They claim that they could
identify the key parameters controlling
the dynamics of the system, using Painlev\'e property (PP)
analysis. They used singular manifold analysis of Weiss,
Tabor and Carnevale (WTC) [Weiss et al., 1983] which was
originally developed for testing necessary conditions for
Painlev\'e property of partial differential equations and
it  is a direct extension of Ablowitz, Ramani and Segur
(ARS) algorithm [Ablowitz et al., 1980] for ordinary
differential equations.  Here we give the correct P-test
results for the model. Numerical evidence given in that
paper is also inconclusive and wrong. Here  we present
results of some extensive numerical calculations and show
the irrelevance of their claims. We also mention some
aspects of the importance of studies on chaos in ecological
models.

\section{The Model}
The three species food chain  model is given by the
following set of ordinary differential equations.

\begin{eqnarray}
{dX\over dt}&=& AX(1-{\frac{X}{K}}) - {\frac{BXY}{(X+D_1)}}
\nonumber\\
 {dY\over dt}&=& -CY + DXY -{\frac{EYZ}{(Y+D_2)}}
\label{1} \\
 {dZ\over dt}&=& -FZ + GYZ, \nonumber
\end{eqnarray}

\noindent
where $A,B,C,D,E,F,G,D_1,D_2$ and $K$ are model parameters
assuming only positive values.  The model describes a prey
population $X$ and two predator populations $Y$ and $Z$.
The predator $Y$ preys on $X$ and the predator $Z$ preys on
$Y$. Here Holling type functional response [Murdoch \&
Oaten, 1975] for the  predator population is  used. Chaos
in a similar model has been studied earlier [Hastings \&
Powell, 1991].

By the following scaling transformations,

\begin{equation}
X \rightarrow Kx, \,\, Y \rightarrow {{KA}\over {B}}y, \,\,
Z \rightarrow {{KA^2}\over {EB}}z, \,\, t \rightarrow
{T\over {A}},
\end{equation}

\noindent
system (\ref{1}) becomes

\begin{eqnarray}
{dx\over dT}&=& x(1-x) - {\frac{xy}{(x+a)}} \nonumber\\
{dy\over dT}&=& -by + cxy - {\frac{yz}{(y+d)}}  \label{2} \\ 
{dz\over dT}&=& -ez + fyz, \nonumber
\end{eqnarray}

\noindent
where $a = {\frac{D_1}{K}}, b={\frac{C}{A}}, c=
{\frac{KD}{A}}, d= {\frac{BD_2}{KA}}, e= {\frac{F}{A}}
\mbox{ and } f={\frac{GK}{B}}$.  This  scaling reduces the
number of parameters in the system from ten to six. The
interesting fact is that, here, $K$ (carrying capacity of the
environment) and $A$ (growth rate of $X$) are absent. Since
this is a continuous system, it can be easily seen that $K$
is just a scaling parameter of the maximum size of
population $X$, which can be arbitrarily fixed at some
value.  By fixing other parameters, $K$ or $A$ may be used
as a control parameter, which is the case with most of the
parameters. Parameter $E$ does not have any effect on
the dynamics of the system.

\section{Painlev\'e Test}
Painlev\'e test (P-test) is a widely used technique to find
the integrability of a dynamical system. It is conjectured
that if the complex time solutions of a system of equations are free
from movable singularities other than poles (Painlev\'e
property), then it is integrable. Several papers and reviews are
available on this technique [Ramani et al., 1989;
Lakshmanan \& Sahadevan, 1993].

We apply here the P-test as devised by WTC [Weiss et al., 1983] on system
(\ref{1}). For testing PP in the case of ordinary differential
equations, one can use ARS algorithm [Ablowitz et al., 1980]
which can be considered as a special case of WTC method.
Here we try to expand the solutions about an arbitrary
singular manifold determined by $\phi(t)=0$, in power
series.

\begin{eqnarray}
X&=& \sum_{j=0}^\infty x_j \phi^{j+\alpha} \nonumber \\
Y&=& \sum_{j=0}^\infty y_j \phi^{j+\beta} \label{4} \\
Z&=& \sum_{j=0}^\infty z_j \phi^{j+n}, \nonumber 
\end{eqnarray}

\noindent
where $x_j, y_j, z_j$ and $\phi$ are time dependent
quantities.  If we replace $\phi$ by $\tau \equiv t-t_0 $,
($t_0$ is position of the singular point), and consider time independent
coefficients, we obtain ARS
P-test, which is easier to do. But WTC method is useful in
finding B\"acklund transformations, Lax Pairs, special
solutions, etc.

Mainly there are three steps in P-test, viz, finding the
dominant behaviour, finding the resonance values and
checking whether there exist arbitrary expansion
coefficients at the resonance values.  Substituting the
$j=0$ terms of (\ref{4}) in (\ref{1}) and equating dominant
terms we obtain, $\alpha=\beta=-1$ and $n=-2$.  Using this
and inserting the full expansion in (\ref{1}) we obtain the
following recursion relations.

\begin{eqnarray}
D_1 [x_{j-2,t}&+&(j-2)x_{j-1} \phi_t] + \sum_{k=0}^j
x_{j-k} [ x_{k-1,t} + x_k (k-1) \phi_t ]\nonumber\\
&-&A(1-\frac{D_1}{K}) \sum_{k=0}^j x_{j-k} x_{k-1} - A D_1
x_{j-2} + B \sum_{k=0}^j x_{j-k} y_{k-1}\nonumber\\
 & +& {{A}\over{K}} \sum_{k=0}^j \sum_{l=0}^k x_{j-k}
x_{k-l} x_l =0, \label{6}\\
 D_2[y_{j-2,t} &+&  y_{j-1} (j-2) \phi_t]
 + \sum_{k=0}^j y_{j-k} [y_{k-1,t} +
 y_k (k-1) \phi_t + C  y_{k-1} + E z_k]    \nonumber\\
&+& C D_2 y_{j-2}  - D D_2 \sum_{k=0}^j
x_{j-k} y_{k-1} - D \sum_{k=0}^j \sum_{l=0}^k x_{j-k}
y_{k-l} y_l  = 0, \label{7}\\ 
z_{j-1,t} &+& (j-2) z_j \phi_t + F z_{j-1} - G \sum_{k=0}^j
y_{j-k} z_k = 0. \label{8}
\end{eqnarray}

	On collecting terms of $x_r, y_r $ and $ z_r$ we
obtain the matrix equation,

\begin{equation}
{\bf M} 
\left(\begin{array}{c} x_r\\ y_r\\z_r \end{array}
\right) = {\bf R_r},
\end{equation}

\noindent
where

\begin{equation}
{\bf M}  =  
\left( \begin{array}{ccc}
(r-2) x_0 \phi_t + {{3A}\over{K}} x_0^2 &0&0\\
-D y_0^2 & (r-2) y_0 \phi_t -2 D x_0 y_0 + E z_0 & E y_0\\
0 & -G z_0& (r-2) \phi_t - G y_0
\end{array} \right),
\end{equation}

\noindent 
and elements of ${\bf R_r}$ are functions of $x_i, y_i, z_i, (i < r) $ and $
\phi $ and their derivatives.  When the determinant of the 
matrix ${\bf M}$ becomes zero, arbitrary coefficients, $x_r , y_r,$ or 
$z_r$ can exist. Those values of $r$ are called resonances.  In
[Rai, 1995] this step is wrong and hence they got wrong
values for the resonances. They have only one resonance 
$r=2$, which is wrong because in a 3-d system there will be
three resonances corresponding to  a generic leading order
behaviour.  The actual resonance values  are  -1 and $ {\frac{1+(KD/A)
\pm \sqrt{K^2D^2/A^2 + 10KD/A +9}}{2}}$.  Resonance value 
-1 corresponds to the arbitrariness of $\phi$. $ r=2$ is a
resonance only when $KD/A=0$, and resonances are not
integers when $KD/A>0$.  Hence the system is nonintegrable.
In the case of weak PP we can allow certain rational
resonance values also but here it is not valid because the
dominant powers are integers [Ramani et al., 1982, Joy \&
Sabir, 1988]. When $D=0$, the equations for $Y$
and $Z$ are decoupled from that of $X$, and there is an integer
resonance $r=2$.  This special case also does
 not have PP because there is only
one positive resonance value; others are negative.

  At the resonance values arbitrary expansion
constants should enter in the expansion for integrability.
But here we can find all the expansion coefficients using
the recursion relations.

For  $j=0$ :

\begin{equation}
 x_0 = {\frac{K \phi_t}{A}},\,\, y_0 = -  {\frac{2 \phi_t}{G}}
 \mbox{ and } z_0 = - {\frac{2 \phi_t^2}{GE}}(1+{\frac{KD}{A}}). 
\label{5}\end{equation}

In [Rai, 1995]  these solutions are given incorrectly and
the expressions they obtained for other $x_j, y_j,$ and $
z_j $ are also  wrong.

  For $j=1$ :

\begin{eqnarray}
x_1&=& - \frac{K}{2 A} \frac{\phi_{tt}}{\phi_t} + {K\over2}
+ {B\over G}\\
 y_1&=& {\frac{\phi_{tt}}{G\phi_t}} - {\frac{[ D_2 +
\frac{2C}{G} - \frac{2 F}{G} (1+ \frac{KD}{A}) +
{{KDD_2}\over{A}} - {{KD}\over{G}} - {{2BD}\over{G^2}} ] }{
( 2 + {{3KD}\over{A}} )}}\\
 z_1&=& ( 1+ {{KD}\over{A}})\, ({{4 \phi_{tt}}\over{GE}}  -
{{2 F \phi_t}\over{GE}}  - {{2 \phi_t y_1}\over{E}} ) 
\end{eqnarray}

	Similarly all $x_j, y_j$ and $ z_j $ for higher
values of $j$ can also be determined using the recursion
relations. For a general solution of the system, we need 3
arbitrary coefficients, which do not exist in this case.
Thus the nonintegrability of the system is confirmed.

According to the expressions in  [Rai, 1995, Eq. (25)], $x_1
$ is determined and their assertion that $X$ is
indeterminate  and hence it is non Painlev\'e type is
misleading. They are getting $y_2$ and $z_2$ in terms of
$x_2$, and claiming that they are arbitrary because $x_2$
is arbitrary!  Moreover they are asserting that the system
is nonintegrable because these coefficients are arbitrary;
but for PP we actually  need arbitrary coefficients at the
resonance values.  They claim that $X$ is the variable
contributing chaotic dynamics to this system which is
otherwise completely integrable [Rai, 1995, last sentence
of the first paragraph in page 540]. Since the system is a
coupled set of 3 differential equations, only one of the
variables cannot be chaotic. When we speak about chaos we
describe the phase trajectories.  It is obvious from basic
dynamical systems  theory that there cannot be chaos in a
2-dimensional system. They have gone to the extent of
saying that since $x_0$ contains the parameters $A$ and
$K$, the indeterministic nature depends on them and they
are the key parameters controlling the dynamics of the
system; and that is their main result!  	Though the
authors did wrong analysis they could give the correct
answer to the question regarding the integrability of the
system in negative. This  happened to be correct because most of the
dynamical systems are nonintegrable.

\section{Numerical Results}
For studying the chaotic behaviour we have plotted the
bifurcation diagram and calculated the maximal Lyapunov
exponents (LE) for various values of the parameters.   
Fig. 1  gives the bifurcation diagram for the
parameter values $B=1,\, C=1,\, D=0.05,\, E=1,\, F=1,\, G=0.05,\, 
D_1=10,\, D_2=10,\, K=50 $, when $A$ is  varied  from 1.0 to 10.0.
These are the same parameter values which has been used in
the paper [Rai, 1995]. To construct the bifurcation diagram
we integrated the system using the above parameter values
and after reaching the attractor (discarded large number of
initial points), we plotted successive maxima (local
maxima) of  the  $Z$ variable, as a function of $A$. The
bifurcation diagram indicates a period doubling route to
chaos in the system. For higher values of $A$,  sequences of
periodic windows and chaos repeat. Though the details are
different, period doubling is observed in each sequence.  
It is clear from the bifurcation diagram that chaotic 
behaviour starts before $A=4.0$ and it is  
confirmed by the Lyapunov exponent calculation. We have
checked our calculations with different initial conditions
also.

We calculated the LE directly from the equations using the
same parameter values.  For calculating the Lyapunov
exponents we have to solve the system alongwith the
corresponding linearised variational system. The method is
available in any book on nonlinear dynamics. (See for
example [Lichtenberg \& Lieberman, 1992]).  Rai [1995] used
Wolf's code for extracting the LE from a time series, by
 taking the $X$ variable as the time series, though Wolf
et al. [1985]  lists a FORTRAN code for calculating the
Lyapunov spectrum from a system of equations also!  Methods of
calculating LE from a time series are not used  when there
are equations known for describing the system, but it is
used when there is only an experimental time series
available. Finding LE from a time series is not at all
reliable and it is always approximate. In the calculation
of invariants such as LE from time series there are a lot
of details to be considered such as the number of data
points, time delay used for reconstruction, the selection
of proper embedding dimension, time used for sampling the
data, etc. There is a large literature available on this
topic of time series analysis. (See for example a review on
this topic by Abarbanel et al. [1993]).  In
Fig. 2 we give maximal Lyapunov exponent of the system as a
function of the parameter $A$ from 3.0 to 10.0, keeping all
other parameters constant.

Our numerical calculations
show that $K$ and $A$ are not the only parameters
determining the dynamics  of the system but most of them
have got some relevance. Detailed studies of such models in general,
will be reported elsewhere. Rai
et al. [1993] claim that they did  not observe chaotic
behaviour by varying other parameters.  

  In Fig. 3 we have plotted $X-Y$
projection of the attractor for $A=4.0$, with all other
parameters are kept the same as that given above. This is
entirely different from the Fig. 2 of [Rai, 1995]. Here we
may note that they have gone wrong somewhere in the
calculation, because in their Fig. 2 of $X-Y$ plot at $A=4$
the $X$ variable reaches values more than the maximum it
can attain which is equal to the value of $K(=50.)$; but it
goes upto $\sim 200$.  It is interesting to note that the
maximal LE we got is one order less than that is given in
[Rai, 1995]. We have done the calculations very accurately
in double precision for different initial conditions and
for different variations and verified our results.

\section{Conclusion}
We have done P-test of the model food chain and found 
it to be nonintegrable. We have explored numerically the
chaotic behaviour of the system by calculating the maximal
Lyapunov exponents, plotting the bifurcation diagram and
phase plots.  We have shown the fallacies of the work and
irrelevance of the claim that P-test can be used to
identify the key parameters determining the dynamics of the
system by Rai [1995]. In  well
known models like Lorenz, Rossler, etc., also P-test
does not give any idea about the key parameters. In any
system, usually all  parameters have  some effect on the
dynamical behaviour. Usually relevant parameters are chosen
by their physical importance. In many systems, dimensions of
the parameter space can be  reduced by proper
transformations, using symmetries, physical considerations,
etc. Singularity structure in the complex time plane is
very much related to the chaotic dynamics of the system,
but to my knowledge no paper  has  appeared in which
P-test is used to identify directly or indirectly 
the key parameters controlling the dynamics of the system.

Selection of biologically realistic parameter values for
the numerical simulation of ecological models is a
difficult problem. Biologically relevant parameter values
can be found by proper identification of the system and its
parameters with natural system for which the model is
applicable.  Studies on chaos such as time series analysis,
prediction techniques, and modelling the system using the
nonlinear dynamical data are useful in this regard.  With
biologically realistic parameters, we can have not only
chaotic behaviour but also simple fixed point behaviour,
limit cycles, etc, depending upon the natural system which
is studied.

Many people misunderstand the importance and the meaning of
chaos in deterministic dynamical systems; they consider the
terminology {\it chaos} of dynamical systems theory for its
literary meaning of total disorder or confusion. We can
study chaos and  even characterize it; moreover there is a
possibility of checking it with the original natural
system. There is a kind of order in chaos that is what we
are interested in and of course the possibility of short
term prediction is always there since it is a deterministic
system [Schaffer 1985]. Modelling of ecological systems and
comparing it with actual ecological data help us to
understand, control, predict, etc., such systems.  It
helps us to understand how the ecology is going to be
affected by various external factors also.

\section*{Acknowledgements}
Author wishes to thank National Board for Higher
Mathematics, DAE, India for financial assistance and Mulugeta
Bekele for critical comments on the manuscript. 

\newpage
\section*{References}
\begin{description}
\item Abarbanel, H. D. I., Brown, R., Sidorowich, J.  J. \&
Tsimring, L. S. [1993] ``The analysis of observed chaotic
data in physical systems'', Rev. Mod. Phys. {\bf 65},
1331-1392.
\item Ablowitz, M. J., Ramani, A. \& Segur, H. [1980] ``A
connection between nonlinear evolution equations and
ordinary differential equations of P-type. I'', J. Math.
Phys.  {\bf 21},  715-721.
\item Gilpin, M. E. [1979] ``Spiral chaos in a
predator-prey model'', American Naturalist {\bf 107},
306-308.
\item  Hastings, A. \&  Powell, T. [1991] ``Chaos in a
three species food chain'', Ecology {\bf 72}, 869-903.
\item  Joy, M. P. \& Sabir, M. [1988] ``Integrability of
two dimensional homogeneous potentials'', J.Phys. A {\bf
21}, 2291-2299.
\item Lakshmanan, M. \& Sahadevan, R. [1993] ``Painlev\'e
analysis, Lie symmetries, and Integrability of coupled
nonlinear oscillators of polynomial type'', Phys. Rep. {\bf
224}, 1-93.
\item  Lichtenberg, A. J. \& Lieberman, M. A. [1992] {\it
Regular and Chaotic dynamics}, (Springer Verlag, New York).
\item May, R. M. [1976] ``Some mathematical models with
very complicated dynamics'', Nature {\bf 261}, 459-467.
\item Murdoch, A. \& Oaten, A. [1975] ``Predation and
Population stability'',  Adv. in Ecological Res. {\bf 9},
1-131.
\item Rai, V. [1995] ``Modelling ecological systems'', Int.
J. Bifurcation and Chaos {\bf 5},  537-543.
\item  Rai, V., Konar, S. \&  Baby, B. V. [1993]
``Painlev\'e property analysis as tool to ascertain the
control parameters of a model food chain'', Phys. Lett. A
{\bf 183}, 76-80.
\item Ramani, A., Dorizzi, B. \& Grammaticos, B. [1982]
``Painlev\'e conjecture revisited'',  Phys. Rev. Lett.
{\bf 49}, 1538-1541.
\item Ramani, A., Grammaticos, B. \& Bountis, T. [1989]
``The Painlev\'e property and singularity analysis of
integrable and nonintegrable systems'', Phys. Rep. {\bf
180}, (1989) 159-245.
\item Schaffer, W. M. [1985] ``Order and Chaos in ecological
systems'', Ecology {\bf 66}, 93-106.
\item  Schaffer, W.  M.  \&  Kot, M. [1985] ``Do strange
attractors govern ecological systems?'', Bioscience {\bf
35}, 342-350.
\item  Weiss, J.,  Tabor, M. \&  Carnevale, C. [1983]
``Painlev\'e property of partial differential equations'',
J.  Math. Phys. {\bf 24},  522-526.
\item Wolf, A., Swift, J. B., Swinney, H. L. \& Vastano, J.
A. [1985] ``Determining Lyapunov exponents from a time
series'', Physics {\bf D16}, 285-317.
\end{description}

\newpage
\section*{Figure Captions}
\begin{description}
\item[Fig. 1]  Bifurcation diagram for the model. Here the
local maxima of $Z$ vs $A$ is plotted. Other parameters are
kept fixed at the values given in the text.
\item[Fig. 2] Maximal Lyapunov exponent vs model parameter
$A$. Other parameter values are the same as that of Fig. 1.
\item[Fig. 3] Projection of the attractor at $A=4.0$ on to
the $X-Y$ plane
\end{description}

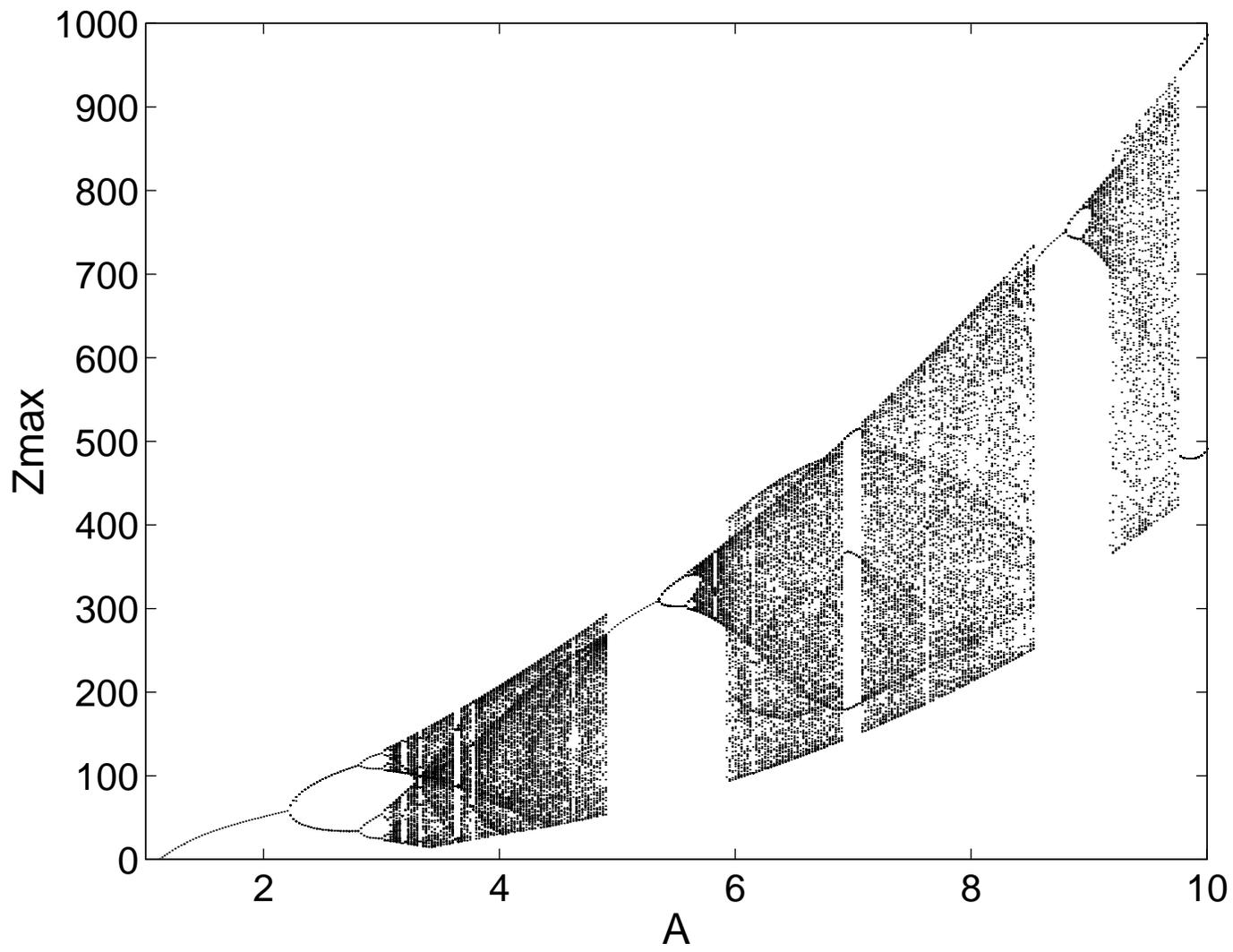
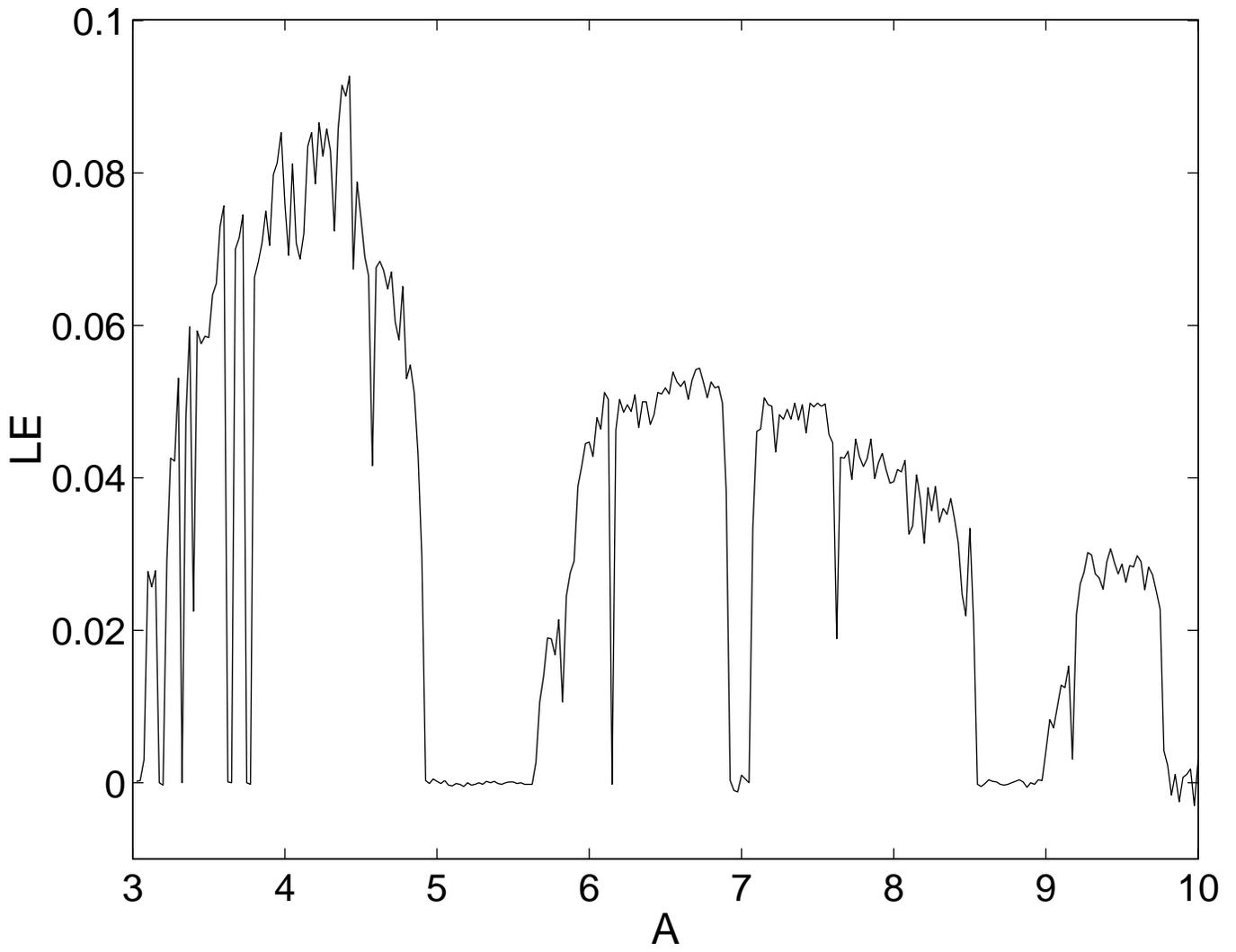
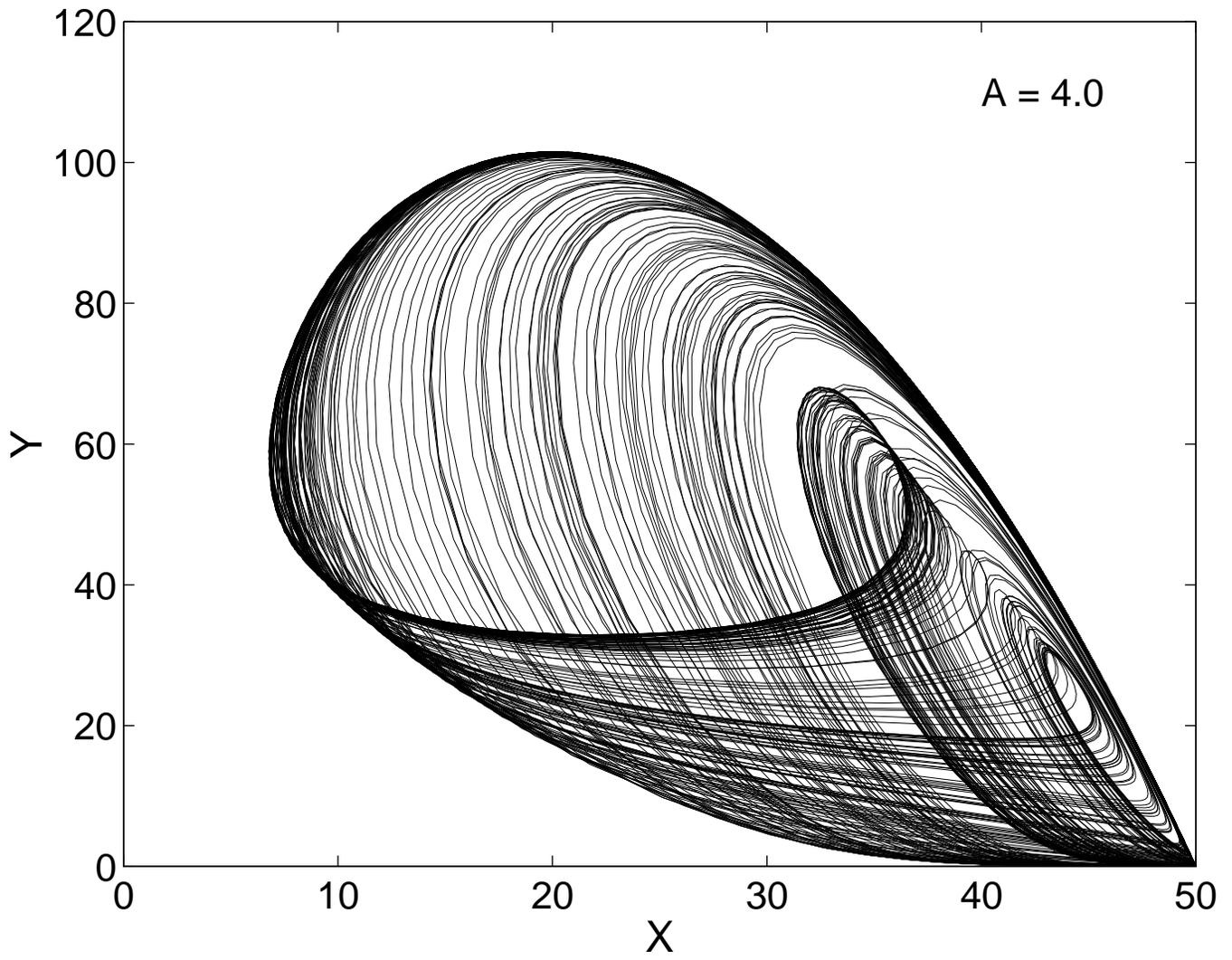
\end{document}